\documentclass[aps,prl,twocolumn,superscriptaddress,showpacs] {revtex4-1}
\usepackage{graphicx}
\usepackage{amssymb}

\begin{document}
\title{New measurement of the scattering cross section of slow neutrons on liquid parahydrogen from neutron transmission}
\author{K.~B.~Grammer}
\email{kgrammer@vols.utk.edu}
\affiliation{University of Tennessee, Knoxville, TN, USA}

\author{R.~Alarcon}
\affiliation{Arizona State University, Tempe, AZ, USA}

\author{L.~Barr\'{o}n-Palos}
\affiliation{Instituto de F\'{i}sica, Universidad Nacional Aut\'{o}noma de M\'{e}xico, Apartado Postal 20-364, M\'{e}xico D. F. 01000, M\'{e}xico}

\author{D.~Blyth}
\affiliation{Arizona State University, Tempe, AZ, USA}

\author{J.~D.~Bowman}
\affiliation{Oak Ridge National Lab, Oak Ridge, TN, USA}

\author{J.~Calarco}
\affiliation{University of New Hampshire, Durham, NH, USA}

\author{C.~Crawford}
\affiliation{University of Kentucky, Lexington, KY, USA}

\author{K.~Craycraft}
\affiliation{University of Tennessee, Knoxville, TN, USA}
\affiliation{University of Kentucky, Lexington, KY, USA}

\author{D.~Evans}
\affiliation{University of Virginia, Charlottesville, VA, USA}

\author{N.~Fomin}
\affiliation{University of Tennessee, Knoxville, TN, USA}

\author{J.~Fry}
\affiliation{Indiana University, Bloomington, IN, USA}

\author{M.~Gericke}
\affiliation{University of Manitoba, Winnipeg, Canada}

\author{R.~C.~Gillis}
\affiliation{Indiana University, Bloomington, IN, USA}

\author{G.~L.~Greene}
\affiliation{University of Tennessee, Knoxville, TN, USA}
\affiliation{Oak Ridge National Lab, Oak Ridge, TN, USA}

\author{J.~Hamblen}
\affiliation{University of Tennessee-Chattanooga, Chattanooga, TN, USA}

\author{C.~Hayes}
\affiliation{University of Tennessee, Knoxville, TN, USA}

\author{S.~Kucuker}
\affiliation{University of Tennessee, Knoxville, TN, USA}

\author{R.~Mahurin}
\affiliation{Middle Tennessee State University, Murfreesboro, TN, USA}
\affiliation{University of Manitoba, Winnipeg, Canada}

\author{M.~Maldonado-Vel\'{a}zquez}
\affiliation{Instituto de F\'{i}sica, Universidad Nacional Aut\'{o}noma de M\'{e}xico, Apartado Postal 20-364, M\'{e}xico D. F. 01000, M\'{e}xico}

\author{E.~Martin}
\affiliation{University of Kentucky, Lexington, KY, USA}

\author{M.~McCrea}
\affiliation{University of Manitoba, Winnipeg, Canada}

\author{P.~E.~Mueller}
\affiliation{Oak Ridge National Lab, Oak Ridge, TN, USA}

\author{M.~Musgrave}
\affiliation{University of Tennessee, Knoxville, TN, USA}

\author{H.~Nann}
\affiliation{Indiana University, Bloomington, IN, USA}

\author{S.~I.~Penttil\"{a}}
\affiliation{Oak Ridge National Lab, Oak Ridge, TN, USA}

\author{W.~M.~Snow}
\affiliation{Indiana University, Bloomington, IN, USA}

\author{Z.~Tang}
\affiliation{Los Alamos National Laboratory, Los Alamos, NM, USA}
\affiliation{Indiana University, Bloomington, IN, USA}

\author{W.~S.~Wilburn}
\affiliation{Los Alamos National Laboratory, Los Alamos, NM, USA}

\date{\today}

\begin{abstract}
Liquid hydrogen is a dense Bose fluid whose equilibrium properties are both calculable from first principles using various theoretical approaches and of interest for the understanding of a wide range of questions in many body physics. Unfortunately, the pair correlation function $g(r)$ inferred from neutron scattering measurements of the differential cross section $d\sigma \over d\Omega$ from different measurements reported in the literature are inconsistent. We have measured the energy dependence of the total cross section and the scattering cross section for slow neutrons with energies between 0.43~meV and 16.1~meV on liquid hydrogen at 15.6~K (which is dominated by the parahydrogen component) using neutron transmission measurements on the hydrogen target of the NPDGamma collaboration at the Spallation Neutron Source at Oak Ridge National Laboratory. The relationship between the neutron transmission measurement we perform and the total cross section is unambiguous, and the energy range accesses length scales where the pair correlation function is rapidly varying.  At 1~meV our measurement is a factor of 3 below the data from previous work.  We present evidence that these previous measurements of the hydrogen cross section, which assumed that the equilibrium value for the ratio of orthohydrogen and parahydrogen has been reached in the target liquid, were in fact contaminated with an extra non-equilibrium component of orthohydrogen. Liquid parahydrogen is also a widely-used neutron moderator medium, and an accurate knowledge of its slow neutron cross section is essential for the design and optimization of intense slow neutron sources. We describe our measurements and compare them with previous work.
\end{abstract}


\maketitle

The physics of liquid hydrogen is of fundamental importance in quantum many body theory.  It is one of the few examples of a dense Bose fluid available for experimental investigation, and it exhibits behavior which interpolates between dense classical liquids and quantum liquids with Bose condensation such as superfluid helium~\cite{Boninsegni2009}. Our ability to understand the physics of this liquid at experimentally-accessible densities and temperatures is important for scientists trying to extrapolate this understanding to predict the properties of the interiors of heavy planets like Jupiter~\cite{Ross1985}. Reliable computational extrapolation to these conditions is thought to require accurate determination of thermodynamic properties of condensed hydrogen at the 1\% level~\cite{Clay2014}.  Metallic hydrogen is also a model system for understanding the metal-insulator transition~\cite{Silvera1980, Li2002, Gregoryanz2003, Tahir2003, Redmer2001, Nellis1999, Ross1996, Weir1996, Mazin1995, Bonev2004}.  Accurate calculations of the properties of liquid hydrogen using theoretical approaches such as Path Integral Monte Carlo (PIMC) and Correlated Density Matrix (CDM) techniques are available~\cite{Bermejo2002, Pavese1996, Bermejo2000, Bermejo1999, Senger1986, Senger1992, Lindenau2002, Gernoth2001} based on well-established input on hydrogen intermolecular potentials such as the Silvera-Goldman potential~\cite{Silvera1978} and the NWB intermolecular potential~\cite{Norman1984}. 

It is therefore disturbing that such a fundamental structural property of liquid hydrogen as the pair correlation function $g®$ (and its Fourier transform partner the static structure factor $S(Q)$) is not yet well determined experimentally. The small electron density makes a measurement using X-rays somewhat difficult. Data on neutron scattering from molecular hydrogen using slow neutrons has been used in the past to help determine $g®$.  In the slow neutron regime the interference scattering from neighboring molecules in the liquid probes a critical region of length scales where the pair correlation function $g®$ is rapidly varying. Unfortunately, neutron scattering experiments which measure the differential cross section $d\sigma \over d\Omega$ and attempt to extract $S(Q)$~\cite{Dawidowski2004, Zoppi1998, Zoppi2002, Celli1999a, Celli2005} are in disagreement. In neutron measurements, the light mass of the hydrogen gives a larger than usual inelastic contribution to the scattering, and large corrections to the scattering data need to be applied in an attempt to extract $g®$. 

\begin{figure}[h!] 
\includegraphics[width=1.0\columnwidth]{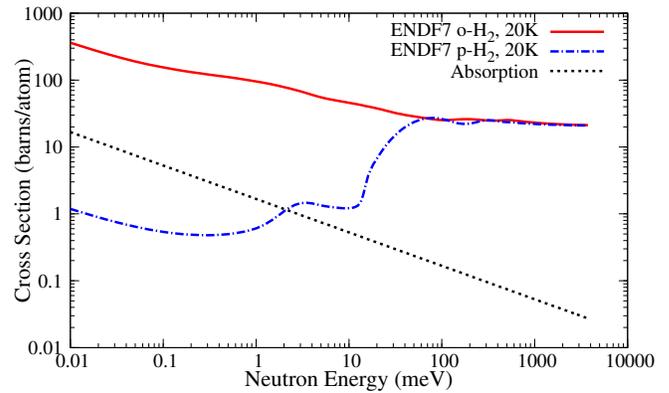}
\caption{Parahydrogen and orthohydrogen scattering cross sections at 20~K from ENDF-VII~\cite{Chadwick2011} and the absorption cross section~\cite{mughabghab2006atlas}.}
\label{fig:cross_sections}
\end{figure}

In this work we present a new measurement of the energy dependence of the total cross section (and, after subtraction of the well-known neutron-proton absorption cross section, the total scattering cross section) in the slow neutron regime using neutrons with energies between 0.43~meV to 16.1~meV in liquid hydrogen at a temperature of 15.6~K$\pm$0.6~K.  This measurement was conducted at the Spallation Neutron Source (SNS) at Oak Ridge National Laboratory using a 16-liter liquid hydrogen target~\cite{Santra2010} operated on the Fundamental Neutron Physics Beamline~\cite{Fomin} (FnPB) by the NPDGamma collaboration. One of the advantages of the sensitive transmission measurement as a function of neutron energy which we report here is that there is no ambiguity in the extraction of the energy dependence of the total cross section $\sigma(E)$. It should therefore be possible to make a more robust comparison of this data with theory. Recall that molecular hydrogen has two spin states, labeled orthohydrogen ($J=\textrm{odd}$) and parahydrogen ($J=\textrm{even}$).  The lowest orthohydrogen state ($J=1$) lies 14.5~meV above the lowest parahydrogen state ($J=0$).  The spin singlet state of the protons in the parahydrogen molecule combined with the measured spin dependence of s-wave neutron-proton scattering amplitudes conspire to greatly suppress the total scattering cross section for neutrons on parahydrogen molecules by more than one order of magnitude relative to that from the hydrogen atom.  The total scattering cross section on orthohydrogen is approximately 50 times higher than on parahydrogen (Fig.~\ref{fig:cross_sections}) because the destructive interference between the atoms is absent. A comparison of our results with previous data~\cite{Seiffert1970c}\cite{Squires1955}\cite{Celli1999a} indicates that the decrease of the total scattering cross section in liquid parahydrogen in the slow neutron regime is much more rapid than previously realized. 

In addition to the usefulness of this new data for extraction of the pair correlation function in liquid parahydrogen, our results are also of immediate practical interest for slow neutron source development. The successful development of intense slow neutron sources combined with the increasing phase space acceptance of neutron optical components has enabled a dramatic expansion of the scientific applications of neutron scattering to encompass many fields in science and technology.  The broad applicability of quantitative information that slow neutron scattering can provide on the internal structure and dynamics of condensed media has motivated the construction of several new neutron scattering facilities over the last decade.  The efficiency of the moderating medium which accepts the relatively high energy neutrons liberated from the nucleus and cools them to the slow neutron energy range below $~25$~meV determines the phase space density of the neutron beams.  New results on physics relevant to the moderation process are therefore of interest to a very broad range of the scientific community.

Many intense neutron sources use liquid hydrogen as a neutron moderator medium.  The near-equality of the neutron and proton mass coupled with the anomalously large s-wave neutron-proton scattering amplitude allow a hydrogen-rich medium to both efficiently lower the incident neutron energy through collisions and also maintain a small neutron mean free path to keep the neutron phase space density high at the source.  In the slow neutron regime, however, the neutron scattering cross section and therefore the mean free path is sensitive to the interference of the scattering amplitudes from neighboring atoms. A neutron that scatters from orthohydrogen will be upscattered and gain 14.5~meV, reducing the slow neutron intensity below 14.5~meV from an orthohydrogen rich moderator.  Consequently, many studies have shown that the slow neutron intensity from a liquid hydrogen moderator can be greatly increased if the molecules are maintained in the parahydrogen molecular state~\cite{Kai2004}\cite{Magan2013}\cite{Ooi2006}.  While absorption ultimately limits the intensity for energies below 2~meV, it is the relative concentrations of orthohydrogen and parahydrogen that is the lever arm available for optimizing the properties of slow neutron moderators. Our new results, which show that the neutron scattering cross section from liquid parahydrogen seems to have been overestimated in previous work by as much as a factor of 3 at an energy of 1 meV, is therefore of immediate interest for the designers of bright slow neutron sources. 

Differences of the orthohydrogen fraction from that corresponding to thermodynamic equilibrium are an obvious culprit for the disagreement among different measurements of neutron-parahydrogen scattering. Given the huge orthohydrogen neutron cross section combined with the well-known difficulty of achieving the proportions of parahydrogen and orthohydrogen in the liquid corresponding to thermodynamic equilibrium, one might be concerned about how closely the nominally liquid parahydrogen samples employed in previous measurements have approached the conditions of thermodynamic equilibrium. Our liquid hydrogen target possessed not only an ortho-para convertor but also a thermosyphon mechanism which forced all of the liquid to pass through the catalyst many times before the neutron transmission measurements were conducted and at a slow but continuous rate during the measurement. We can take the functional form of our measured neutron cross section as a function of neutron energy on liquid parahydrogen and reproduce previous cross section data by adding to it an extra component of orthohydrogen scattering using the measured neutron energy dependence of scattering on orthohydrogen. This result strongly suggests to us that the nominally liquid parahydrogen samples used in previous total cross section measurements in fact possessed higher residual orthohydrogen contamination than expected based on thermodynamic equilibrium. 

\begin{figure}[h!] 
\includegraphics[width=.45\textwidth]{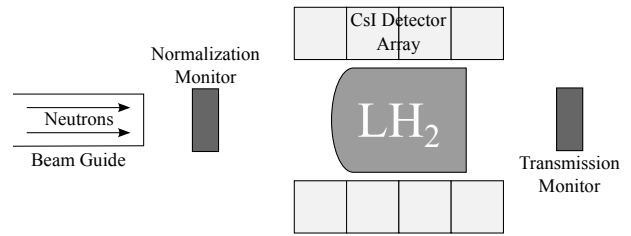}
\caption{Experimental setup showing the cesium iodide detector array, liquid hydrogen target, and beam monitors}
\label{fig:setup}
\end{figure}

Figure \ref{fig:setup} shows the experimental setup for the transmission measurement on the FnPB beamline at the SNS.  The cross sectional area of the neutron beam is 12$\times$10~cm$^2$ at the exit of the neutron guide.  Neutrons then pass through the normalization monitor, a multi-wire proportional counter~\cite{mccreamonitors} with a gas mixture of $^3$He (15.1~Torr) and N$_2$ (750~Torr) located 15.24~m $\pm$ 0.12~m from the moderator, and are incident on the 16 liter liquid hydrogen target~\cite{Santra2010} centered 17.6~m from the moderator.  Roughly 60\% of the neutron beam is captured on the hydrogen in the target with the 2.2~MeV capture gammas relevant to the NPDGamma experiment detected by 48 cesium iodide crystals.  The effective length of liquid hydrogen covered by the neutron beam cross sectional area including beam divergence is 30.065~cm $\pm$ 0.005~cm when cold.  The transmitted neutrons exit the downstream end of the target vessel through a 2.5~cm diameter aperture in the $^{6}$Li-rich neutron absorber surrounding the target.  The transmitted neutron intensity is measured in a $^3$He plate ion chamber~\cite{Szymanski1994} located 3.44~m $\pm$ 0.02~m from the normalization monitor.  The charge produced in the monitors is amplified by current-voltage amplifiers~\cite{ArtifexEngineering2010} with a 10~kHz bandwidth.

The data acquisition system records data in 0.4~ms increments.  In order to avoid contamination from overlapping neutron pulses and to increase the dynamic range of neutron energies for the transmission measurement, data were taken with the two beamline choppers parked open while the SNS was operating at 10~Hz duty cycle rather than the normal 60~Hz.
\begin{figure}[h!] 
\includegraphics[width=0.45\textwidth]{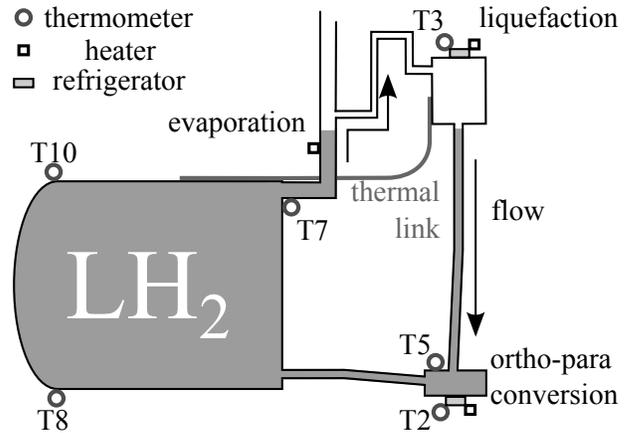}
\caption{Diagram of circulation loop inside the hydrogen target system.  Evaporated hydrogen is re-condensed and is forced to flow through the ortho-para convertor (OPC) at a rate of a few millimoles per second.  T3, T7, T8, and T10 determine the liquid hydrogen bulk temperature.  T2 and T5 determine the temperature of the catalyst in the OPC.}
\label{fig:target}
\end{figure}

\begin{figure}[h!] 
\includegraphics[width=1.0\columnwidth]{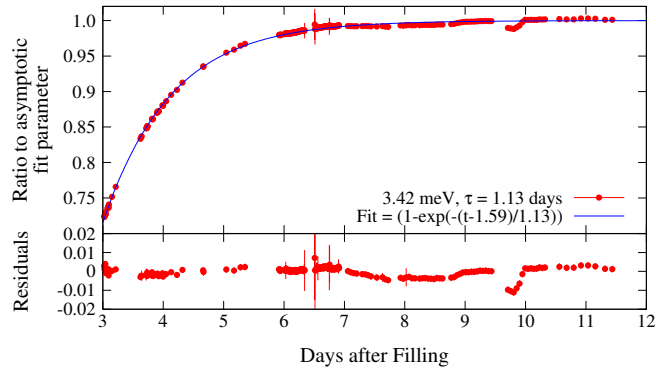}
\caption{Observed ortho-para conversion over time as a fraction of the asymptotic limit for 3.42~meV neutrons shortly after filling the target, with a time constant of approximately 1 day.  Residuals from the exponential fit are shown at bottom.}
\label{fig:transmission}
\end{figure}

The target vessel is initially filled with hydrogen gas, corresponding to 3 orthohydrogen molecules per parahydrogen molecule from equipartition.  The equilibrium parahydrogen concentration increases with decreasing temperature~\cite{Dennison1927}.  The slow natural conversion to parahydrogen is accelerated by circulation of the liquid through 150~mL of hydrous iron (III) oxide~\cite{catalyst} 30 - 50 mesh powder catalyst in the ortho-para converter (OPC)~\citep{Barron-Palos2011} in the NPDGamma target loop~(Fig.~\ref{fig:target}).  The neutron transmission increases with time as hydrogen circulates through the catalyst until a steady-state condition is reached.  Fitting the transmission for 3.42~meV neutrons to an exponential as a function of time (Fig.~\ref{fig:transmission}) indicates that the parahydrogen concentration in the main target vessel approaches saturation.  This exponential approach to the steady-state condition implies that the conversion is dominated by the first order processes in the OPC as liquid hydrogen circulates through the catalyst.

The conversion process shown in figure \ref{fig:transmission} has reached steady-state, where the parahydrogen concentration is near the thermal equilibrium value defined by the temperature of the OPC.  The average temperature of the OPC was 15.4~K$\pm$0.5~K, which corresponds to a thermal equilibrium parahydrogen concentration of 0.99985.  A small amount of para-to-ortho conversion may take place in the liquid in the main vessel, on the walls of the vessel, or the walls of the circulation loop that prevents reaching absolute thermal equilibrium.  Since para-to-ortho conversion is known to be a very slow process it is not expected to limit the ortho-para ratio in the liquid hydrogen, and the liquid hydrogen is expected to be in thermal equilibrium with the catalyst.  However, it was not possible to independently confirm the parahydrogen concentration in this system.

Two different measurements were required in order to measure the empty target and full target transmissions.  The full target measurement was performed over 8 hours with the target vessel at 15.6~K after the target had been in steady-state operation for 4 weeks, which corresponds to 30 conversion time constants.  The empty target measurement was performed 2 weeks later with the target vessel at 16.3~K in order to cancel the temperature dependence of scattering from the aluminum target vessel.  Between these two measurements, the moderator viewed by the beamline was emptied and refilled with fresh liquid hydrogen, which led to a small change in the moderated neutron spectrum between the two measurements.  A neutron energy dependent correction was applied to the transmission monitor signals using the ratio of the normalization monitor signals to account for this systematic effect.  For each neutron pulse, the signals in each monitor are normalized to the per-pulse beam power by integrating the normalization monitor over peak signal range.

\begin{figure}[h!] 
\includegraphics[width=1.0\columnwidth]{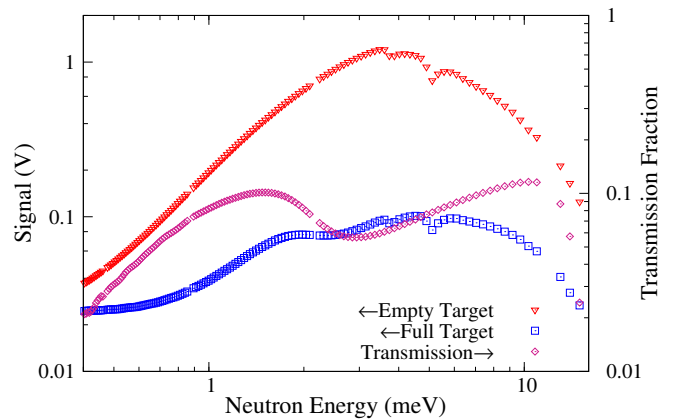}
\caption{(color online) Transmission monitor signals (left axis) for empty (triangles) and hydrogen-filled (squares) aluminum target vessel.  Dips in the spectra are at the aluminum Bragg edges.  Transmission ratio (right axis, diamonds) depicts no transmission for energies above 14.5~meV spin-flip transition.}
\label{fig:full_and_empty}
\end{figure}

There is not a direct correspondence between time of flight bins in each monitor due to time of flight broadening.  The normalization monitor signal is fit to a cubic spline in order to interpolate for spectrum normalization.  The sharp dips in the pulse shapes in figure \ref{fig:full_and_empty} are due to Bragg scattering on aluminum windows along the path of the neutron beam.  These dips are visible at neutron energies of 4.98~meV and 3.74~meV, corresponding to the aluminum (200) and (111) Bragg planes~\cite{Wyckoff1963}, respectively.  The time of flight of the Bragg edges for the normalization monitor is used to determine the distance from the moderator and to convert each time of flight bin to neutron energy.  The uncertainty in time due to these Bragg edges is 0.16~ms, which determines the uncertainty in the normalization monitor position.  The target-full spectrum indicates no measurable neutron flux for energies above 14.5~meV (Fig.~\ref{fig:full_and_empty}).  This is the minimum energy necessary for the $J=0 \to 1$ spin-flip transition, meaning neutrons with energies above this threshold are scattered out of the beam rather than transmitted through the target.  The data also contain a 240~Hz noise component with an amplitude of a few millivolts.  The amplitude is diminished by averaging pulses over the measurement period and is only visible for small signals.  The transmission monitor signals at long wavelengths are fit to a sinusoidal function corresponding to the 240~Hz noise.  The sinusoidal function is subtracted before extracting the transmission.  After correcting for the pedestal, 240~Hz noise, and moderator spectrum, the final corrected transmission (Fig.~\ref{fig:full_and_empty}) is given by
\begin{equation}
T(\lambda) = \frac{S_{\textrm{trans,full}}(\lambda)}{S_{\textrm{trans,empty}}(\lambda)}\frac{S_{\textrm{norm,empty}}(\lambda)}{S_{\textrm{norm,full}}(\lambda)}\frac{g_{\textrm{norm}}}{g_{\textrm{trans}}},
\label{eq:transmission_majesty}
\end{equation}
where the $S$ values are monitor signals and $g$ are monitor gain adjustment factors.

The contamination of the transmission signal by non-forward small angle neutron scattering in our geometry was estimated to be less than 0.1\% in MCNPX~\cite{Pelowitz2011} using the ENDF-VII thermal cross sections~\cite{Chadwick2011}.  The total cross section can then be written as:

\begin{eqnarray}
\label{eq:sigma}
\sigma_{\textrm{total}}(\lambda)&=&\frac{-\log[T(\lambda)]}{nl} \\
&=& \sigma_{\textrm{abs}}(\lambda) + \sigma_{\textrm{scatter}}(\lambda) \nonumber \\
&=& \sigma_{\textrm{abs}}(\lambda) + f\times \sigma_{\textrm{para}} + (1-f)\sigma_{\textrm{ortho}}\nonumber, 
\end{eqnarray}
where $n$ is the number density, $l$ is the hydrogen length, $f$ is the parahydrogen fraction, $\sigma_{\textrm{abs}}=0.3326\pm0.0007$~barns at 2200m/s~\cite{mughabghab2006atlas}, $\sigma_{\textrm{scatter}}$ is the total scattering cross section, $\sigma_{\textrm{ortho}}$ is the orthohydrogen scattering cross section, and $\sigma_{\textrm{para}}$ is the parahydrogen scattering cross section.

The diode temperature sensors have an accuracy of 0.5~K and upward drift due to radiation damage is not worse than 0.3~K, providing a total uncertainty on the temperature of 0.6~K.  The density of the liquid hydrogen in our target is determined from a fit to data compilations of the density of liquid hydrogen as a function of temperature from many sources \cite{Leachman2009}\cite{Souers1986}\cite{McCarty1981}.  The transmission data include several instrumental effects such as the monitor efficiency, the monitor dead layer, and monitor linearity.  These effects all cancel in equation~\ref{eq:transmission_majesty} as long as the monitors and preamplifiers are linear and the aluminum components of the experiment were maintained at the same temperature.  The linearity of the transmission monitor was determined from a scan of the bias voltage in order to reduce volume recombination effects in the chambers, with a resulting uncertainty of 0.15\% for each monitor.  Controlled current injection was used to measure the linearity of preamplifiers and the gain shift, which are 0.01\% and 0.1\% respectively.

\begin{table}
\begin{center}
\caption{Main uncertainties in the total cross section at 1.92~meV}
\label{tab:error_budget}
\begin{tabular}{| l | c |}
\hline
Source & Uncertainty \\
\hline
Neutrons & 0.02\% \\
Time of Flight & 0.61\% \\
Monitor Gains & 0.06\% \\
Monitor Linearity & 0.12\% \\
Target Length & 0.007\% \\
Liquid Density Fit & 0.5\% \\
Temperature & 0.71\% \\ 
\hline
Total & 1.07\% \\
\hline
\end{tabular}
\end{center}
\end{table}

\begin{figure}[h!] 
\includegraphics[width=1.0\columnwidth]{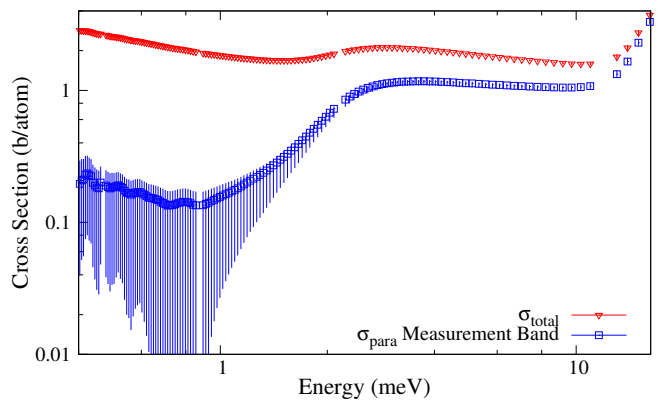}
\caption{(color online) Total cross section from this work in barns/atom (triangles), parahydrogen scattering cross section (squares).  The upper error bar on the parahydrogen cross section comes from table~\ref{tab:error_budget} and the lower error bar is given by the upper limit on the orthohydrogen contamination.}
\label{fig:sigma_all_10hz_log}
\end{figure}

\begin{figure}[h!] 
\includegraphics[width=1.0\columnwidth]{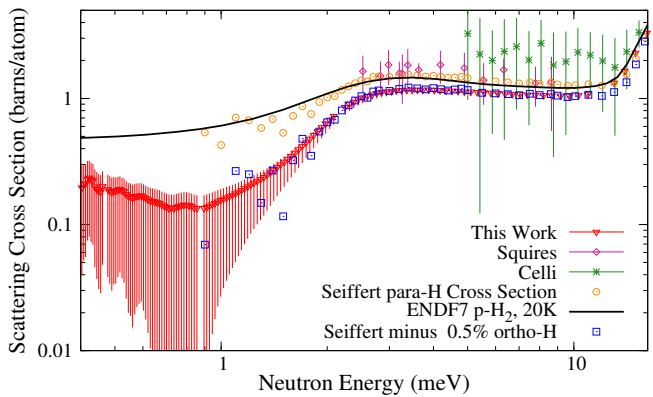}
\caption{(color online) The scattering cross section extracted in this work (triangles), Squires~\cite{Squires1955} (diamonds), Celli~\cite{Celli1999a} (stars, some points omitted), Seiffert~\cite{Seiffert1970c} (circles), ENDF-VII (black), and subtraction of a 0.5\% admixture of orthohydrogen from Seiffert (squares).}
\label{fig:sigma_scatter}
\end{figure}

We have determined the total cross section for liquid hydrogen at 15.6~K from approximately 0.43~meV to 16.1~meV with an uncertainty of approximately 1\%, or 0.02 barn/atom over the majority of the measurement range (Fig.~\ref{fig:sigma_all_10hz_log}).  Because the absorption cross section is well known, we are also able to determine a measurement band for the parahydrogen scattering cross section at these energies.  This measurement band is much smaller than the values previously reported in the literature~(Fig.~\ref{fig:sigma_scatter})~\cite{Seiffert1970c}\cite{Squires1955}\cite{Celli1999a}, with the Seiffert cross section predicting a transmission for our apparatus that is 2\% less than was measured at the lowest energies.  Furthermore, we can set an upper limit on the orthohydrogen concentration in our apparatus by attributing all of the scattering at 0.8~meV to orthohydrogen, which results in an upper limit on the orthohydrogen concentration of 0.0015 using the ENDF-VII orthohydrogen cross section.  At the lowest energies, we cannot distinguish the parahydrogen cross section from zero, however, we can put a band on the parahydrogen cross section at higher energies.  The central value corresponds to the parahydrogen concentration given thermodynamic equilibrium in the OPC, 0.99985.  The upper error bar on this central value is determined by the uncertainties presented in table \ref{tab:error_budget} and is dominated by the temperature and the time of flight.  The lower error bar is determined by the orthohydrogen upper limit and is determined by the orthohydrogen cross section from ENDF-VII scaled by a factor of 0.0015.

The measurement of the parahydrogen scattering cross section is very sensitive to the orthohydrogen fraction in the target volume because the orthohydrogen cross section is approximately a factor of 50 greater than for parahydrogen.  The parahydrogen scattering cross section from this work along with the Seiffert~\cite{Seiffert1970c} data and the ENDF-VII parahydrogen kernel evaluated at 20~K~\cite{Chadwick2011} are compared in Fig.~\ref{fig:sigma_scatter}.  The significant difference in magnitude suggests the presence of unaccounted for orthohydrogen contamination in previous experiments.  Subtraction of an admixture of 0.5\% orthohydrogen from Seiffert data brings both results into agreement.

The Squires measurement~\cite{Squires1955} was performed using a gas mixture with a parahydrogen concentration of 0.9979, which was independently measured using thermal conductivity.  The Seiffert~\cite{Seiffert1970c} and Celli~\cite{Celli1999a} measurements were both performed using liquid hydrogen in the presence of a catalyst; however, neither experiment independently determined the orthohydrogen concentration but rather inferred that it was either negligible, in the case of Seiffert, or at thermal equilibrium, in the case of Celli.  We therefore treat both the Seiffert and Celli measurements as upper limits.  We conclude that our target system must have less orthohydrogen contamination than these previous two measurements because our observed total cross section is lower.  Of these three measurements in the literature and the measurement in this work, we believe that our measurement has the lowest orthohydrogen contamination and that it provides the most accurate measurement of the liquid parahydrogen scattering cross section.

These results have important implications for the design of slow neutron sources.  Recent simulation work conducted for the European Spallation Source project~\cite{Magan2013}, indicates increased source intensity from liquid parahydrogen neutron moderators incorporated into a realistic target-moderator geometry.  Measurements at J-PARC~\cite{Kai2004} and LANSCE~\cite{Ooi2006} also show that the moderator intensity for neutrons below 14.5~meV are highly dependent on the ortho/para ratio.  Our work shows that the parahydrogen cross section has been previously overestimated throughout the slow neutron regime of interest.  This overestimate reaches a factor of 3 at a neutron energy of 1~meV.  The potential for increased slow neutron source intensity from liquid parahydrogen moderators is therefore greater than previously realized and impacts the optimal geometry of slow neutron moderators.  In order to be able to take full advantage of this potential, however, it would be necessary to maintain the liquid in the parahydrogen state in the presence of the intense radiation environment accompanying an intense neutron source~\cite{Iverson2003}.  Liquid hydrogen target designs which employ active circulation of the hydrogen through a catalyst coupled with dedicated measurements of the parahydrogen fraction from a liquid hydrogen moderator operated in an intense radiation environment are needed to confirm this potential and demonstrate that it can be realized at an intense neutron source.

We would like to thank Erik Iverson, Phillip Ferguson, Kenneth Herwig, and Franz Gallmeier for productive discussions and encouragement for this experiment as well as Michael Mendenhall for thoughtful observations.  We also thank the management and staff of the Spallation Neutron Source for adapting our measurement to the busy beam delivery schedule.  We gratefully acknowledge the support of the U.S. Department of Energy Office of Nuclear Physics (including Grant No. DE-FG02-03ER41258), the National Science Foundation (including Grant No. PHY-1068712), PAPIIT-UNAM (Grant No. IN111913), and the Indiana University Center for Spacetime Symmetries.

\bibliographystyle{apsrev4-1}
\bibliography{papers_writeups-hydrogen_sigma_edit.bib}

\end{document}